# Detecting and Analyzing Mobility Hotspots using Surface Networks[1]

Yujie Hu[1]*, Harvey J. Miller[2], and Xiang Li[1]

[1]Key Laboratory of Geographical Information Science, Ministry of Education, East China Normal University, Shanghai, China

[2]Department of Geography, The Ohio State University

Corresponding author: Yujie Hu

**Abstract.** Capabilities for collecting and storing data on mobile objects have increased dramatically over the past few decades. A persistent difficulty is summarizing large collections of mobile objects. This paper develops methods for extracting and analyzing *hotspots* or locations with relatively high levels of mobility activity. We use kernel density estimation (KDE) to convert a large collection of mobile objects into a smooth, continuous surface. We then develop a topological algorithm to extract critical geometric features of the surface; these include *critical points* (peaks, pits and passes) and *critical lines* (ridgelines and course-lines). We connect the peaks and corresponding ridgelines to produce a surface network that summarizes the topological structure of the surface. We apply graph theoretic indices to analytically characterize the surface and its changes over time. To illustrate our approach, we apply the techniques to taxi cab data collected in Shanghai, China. We find increases in the complexity of the hotspot spatial distribution during normal activity hours in the late morning, afternoon and evening and a spike in the connectivity of the hotspot spatial distribution in the morning as taxis concentrate on servicing travel to work. These results match with scientific and anecdotal knowledge about human activity patterns in the study area.

**Keywords:** Mobile objects; hotspots; kernel density estimation; surface feature extraction; graph theory

## 1. Introduction

Capabilities for collecting data on mobile objects such as vehicles have increased remarkably over the past few decades. These data are valuable for many scientific, planning and management applications, including navigation, road pricing, traffic flow analysis and fleet management (Jagoe 2002). A persistent issue is how to summarize collections of mobile objects. The data for a single mobile object consists of a large number of sample points recorded by GPS or other positioning devices, often at the rate of a new sample point per 1-2 seconds. The complexity of extracting meaningful

---

properties from these sample points increases dramatically as the number of mobile objects increases. This has motivated the development of methods for *mobility mining* (Andrienko and Andrienko 2008). Mobility mining techniques include identifying *clusters* or groups of similar trajectories, *frequency patterns* reflecting repeatedly followed paths or subpaths and *classifications* based on geometry or semantics (Han et al. 2010). A useful concept in mobility cluster analysis is *hot spots* or locations and times displaying a relatively high level of mobility activity (Brimicombe 2005). Hot spots can indicate travel demand patterns, social dynamics (such as areas within a city with lively night life) or unusual events (such as abnormal movements due to unplanned infrastructure disruptions or special events such as concerts).

This paper develops methods for identifying and analyzing hot spots in collections of mobile objects. The basic idea is to summarize the locations of the mobile objects over some interval of time as a surface, simplify the surface by extracting critical features, generate a network representing the feature topology and analyze the network using graph theoretical measures. We use kernel density estimation (KDE) to convert the mobile object locations into a surface, develop algorithms to extract *critical points* (peaks, pits and passes) and *critical lines* (ridgelines and course-lines) and connect the peaks and ridgelines to produce a surface network. A wide array of graph theoretic measures can be used to analyze the surface network as well as compare surface networks representing the mobile objects over a sequence of time. To illustrate our approach, we apply the techniques to taxi cab data collected in Shanghai, China.

The next section of this paper provides some background. The next section presents the methodology; this includes KDE, techniques for extracting critical features and generating a corresponding surface network, and a set of graph theoretic measures for quantitatively summarizing the network. Section 4 illustrates the methods using mobile objects data collected in Shanghai, China. Section 5 concludes the paper with a discussion of the method's strengths and weaknesses, as well as directions for additional research.

## 2. Background

The concept of *hotspots* or locations with relatively high levels of an attribute or activity originated in the field of criminology (Sherman 1989, Sherman 1995, Weisburd and Green 1995, Wen et al. 2010). In this domain, hot spots are areas where a greater number of criminal or disorderly events happen relative to other locations (Eck et al. 2005). Identifying and analysis of crime hotspots are useful for policy analysis and resource allocation, for example, identifying whether different policing strategies are effective and identifying areas where attention and resources should be concentrated.

The hotspot concept has been extended to many different fields including transportation and epidemiology. Transportation applications include identifying traffic accident hotspots and understanding travel demand dynamics (Yamada and Thill 2004,

Anderson 2007, Erdogan et al. 2008). Traffic accident hotspots represent areas with higher probabilities of accidents; identifying these hotspots is important for the appropriate allocation of resources for safety improvements (Anderson 2009). With respect to travel demand dynamics, hotspots are areas with relatively high levels of mobility activity. Identifying these hotspots and their spatial patterns can provide insights that allow better allocation of transportation infrastructure and especially services that can be configured dynamically, such as congestion management and public transit schedules (Kaysi et al. 2003, Chu and Chapleau 2010). Applications in epidemiology concentrate on the identification of high prevalence areas of diseases such as cancer, diabetes and dengue (Kulldorff et al. 1998, Green et al. 2003, Joseph and Laurie 2005, Jeefoo et al. 2010). Identifying hotspots for a disease and their spatial distributions is important for public health surveillance, allowing better understanding of the dynamics of disease diffusion at locations where surveillance, control, prevention and other related services should be provided, and better exploring of the reasons or affecting/causing factors associated with the disease (Tiwari et al. 2006, Jeefoo et al. 2010).

There are two major approaches for extracting hotspots. One approach is cluster analysis which addresses the problem of finding subsets of interest in a database (Anderberg 1973, Kaufman and Rousseeuw 1990). Clusters are required to be *homogeneous* (meaning that entities within the same cluster should resemble one another) and/or *well-separated* (meaning that entities in different clusters should differ one from the other; Hansen and Jaumard 1997). With respect to hotspot identification, cluster analysis seeks to segment space into regions with similarly high levels of a spatial attribute relative to neighboring locations (Brimicombe 2005).

A second approach to hotspots detection is to identify locally high levels of an attribute occurrences using spatially exhaustive search (Brimicombe 2005). An effective search method is *kernel density estimation* or KDE (Silverman 1978, Silverman 1981, Silverman 1986, Kelsall and Diggle 1995). KDE generates a surface summarizing the spatial distribution of a point pattern, and high points on the surface (typically defined by stating a threshold value) are identified as hotspots. KDE has two characteristics that make it well suited for hotspot identification (Chainey and Ratcliffe 2005). First, KDE has the ability to determine arbitrary spatial regions for each hotspot. Second, the spatial unit of analysis can be defined flexibly (Anderson 2009). Consequently, KDE has been applied in the analysis of crime, traffic accidents, spatial market analysis and the spread of disease (Donthu and Rust 1989, Bithell 1990, Rushton and Lolonis 1996, Rushton 1998, Gahegan 2003, Anderson 2009, Demšar and Virrantaus 2010, Wen et al. 2010).

The surface generated through KDE is useful as a visual summary of the underlying point patterns and can be subjected to quantitative analysis (De Floriani et al. 1996,, Sadahiro 2001, Sadahiro and Masui 2004, Kobayashi et al. 2010). Quantitative surface measures can include altitude, slope and aspect at any given location. These measures can be compared among surfaces through *agreement indices*. Agreement indices compare the values at sample locations on the surface and measure whether the values concur at corresponding locations. A regular mesh of these comparison points can be used to generate a surface that summarizes the agreement among surfaces (Sadahiro and Masui 2004).

Other surface analysis measures attempt to summarize morphology. *Surface networks* are simpler representations of surfaces based on topological graphs. A surface network consists of critical points and critical lines. Critical points defined are surface locations that contain sufficient information to characterize local surface feature such as peaks (local highpoints), pits (local low points and passes (saddle-like features connecting peaks and separating pits). Surface networks comprised of critical lines connecting critical points are an effective representation of the topographic structure of the entire surface (Warntz 1966, Warntz 1975, Heil and Brych 1978, Okabe and Masuda 1984, Sadahiro 2001, Rana 2004, Wolf 2004). Surface networks are also efficient data structures for accessing surface data and allow extraction of useful physical parameters such as traversals and drainages (Pfaltz 1976, Kreveld et al. 1998, Biasotti et al. 2008, Jeong et al. 2014). A surface network can also indicate the underlying complexity of a spatial distribution. Okabe and Masuda (1984) interpolate surfaces from population data and classify 245 cities into a small number of classes based on the size and complexity of the resulting networks as measures of the underlying social morphology of the cities.

Surface networks lend themselves to quantitative measurement and comparison. Sadahiro and Masui (2004) develop measures to compare these features among surfaces. These measures include α-peak, α-pit and β-monotonic line, where the parameters refer to the minimum ratio of all surfaces that share that feature within a local neighborhood. Sadahiro (2001) develops more explicitly temporal methods that identify events corresponding to changes in critical points such as *generation*, *disappearance*, *movement* and *switch*.

Since surface networks are graphs, it is possible to summarize and compare their topology using *graph indices*. There is a well-established literature in graph theory and related indices dating back over 50 years (Shimbel 1953, Kansky and Danscoine 1989). Graph indices are applied across a wide range of fields (in particular, transportation) and remain an area of active investigation (Xie and Levinson 2007, Derrible and Kennedy 2009, Rodrigue et al. 2009). We will describe graph indices more fully in the subsequent section.

## 3. Methodology

Figure 1 describes the methodology workflow. For each time period $t_i$ (this may be an instant or an interval), we have a collection of trajectory points recorded during that period. Kernel Density Estimation (KDE) generates a surface that summarizes the point distribution. Algorithms extract critical surface features and generate a surface network based on these features. Graph indices summarize and compare the spatial distribution of the points across different time periods. We describe each of these steps in more detail below.

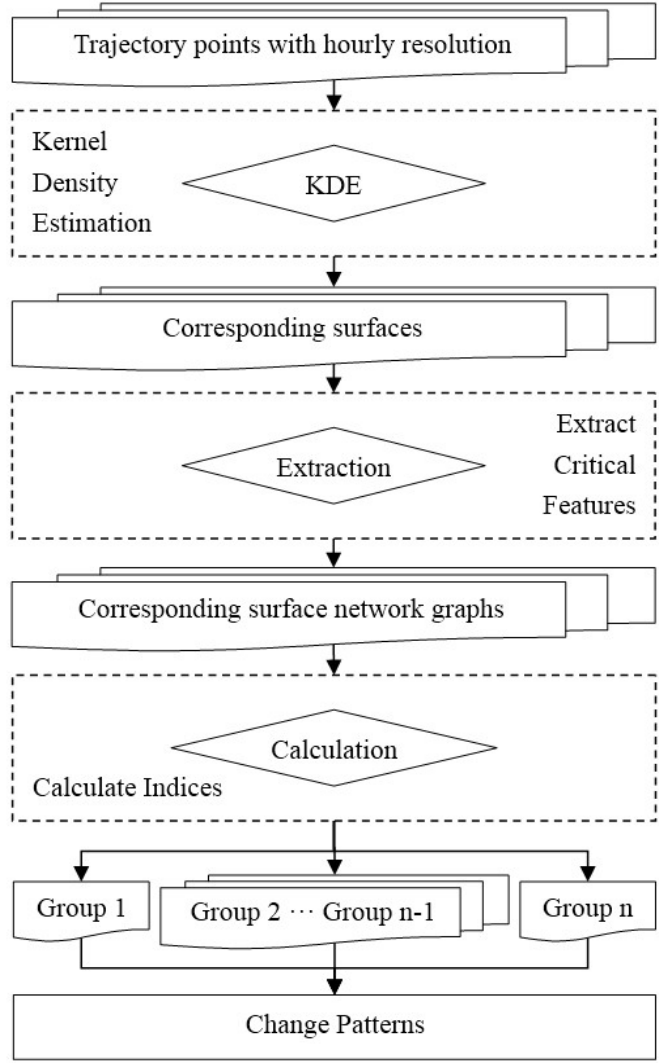

Figure 1. Workflow of the proposed method

## 3.1 Kernel density estimation

The Kernel Density Estimation (KDE) method is:

$$f(x) = \frac{1}{nh^2} \sum_{i=1}^{n} K\left\{\frac{(x-x_i)}{h}\right\} \qquad (1)$$

where $n$ is the number of data points, $h$ is the bandwidth, $K$ is kernel function, $x$ is a vector of coordinates indicating the location where the density is being estimated, and $X_i$ is a series of vectors of corresponding location coordinate of each observation $i$. Figure 2 illustrates the basic principles. The method first divides the entire study region into predetermined number of grid cells. It then delineates a circular area with a radius of the bandwidth $h$ around each data point. The kernel function weights the interpolation, ranging from 1 (at the location of the data point being evaluated) to 0 (at the location of the circular area boundary) at each sample data point $(x_i, y_i)$. In this way, the KDE method places a quadratic surface indicating the probability density over each sample data point. After calculating kernel densities at each data point, we convert the densities into a raster surface by estimating the density value at each raster cell based on density values of all kernel surfaces that intersect with the center of the cell.

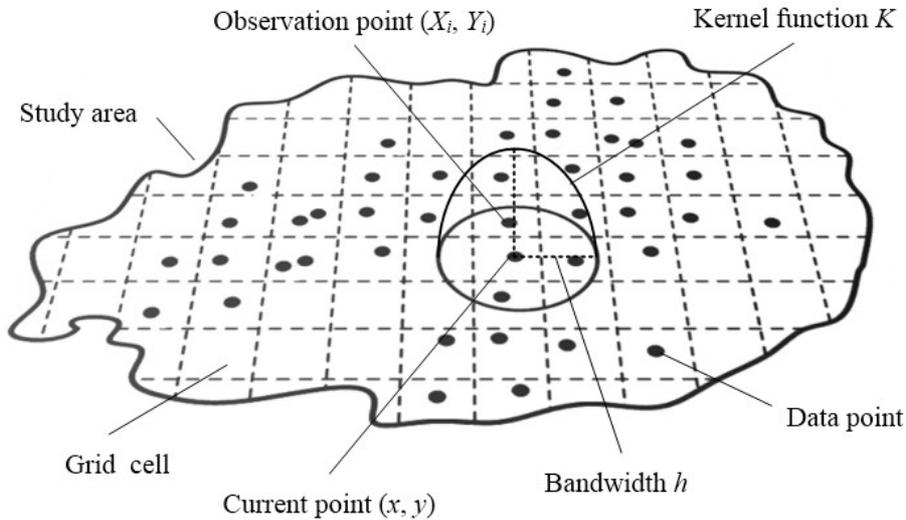

Figure 2. KDE method (based on Anderson 2009)

KDE requires two parameters, the bandwidth and output grid cell size, with the bandwidth $h$ being especially critical (Anderson 2009; Bailey and Gatrell 1995; Silverman 1986). At small bandwidths, we would detect small, spiky hotspots but with higher probability of false positives. Noise decreases as bandwidth increases, but we bias detection towards smoother and larger hotspots and fail to detect smaller hotspots, raising the probability of false negatives. Several methods for determining the appropriate bandwidth setting are available, including data-based selection (Sheather and Jones 1991), cross-validation (Brunsdon 1995) and distance-based approaches (Fotheringham et al. 2000). In this paper, we experimented with several different bandwidth values and finally chose a value of 600 meters as the desired bandwidth in order to detect significant and sharp hotspots as well as minimize the occurrences of false positives hotspots. We will discuss this in more detail in section 4.

## 3.2. Extraction of Critical Features

Mathematically, a surface (denoted by $S$) can be represented by a single-valued function $z = f(x, y)$ in the three-dimensional Cartesian space, where $z$ means the height value associated with each point $(x, y)$ in the two dimensional plane. A common assumption for feature extraction is that the surface is smooth and differentiable at any point $(x, y)$ in $S$ (Okabe and Masuda 1984, Sadahiro 2001, Sadahiro and Masui 2004), a property that KDE-generated surfaces meet. Given this surface, we now describe how to extract its critical points.

**Critical points.** Rana (2004) defines critical points as characteristic local surface features that contain sufficient information to reconstruct the surface, thus taking away the need to store the entire surface. It is commonly assumed that there exists a local surface (denoted by $S_p$) around each point $(x, y)$ in $S$, and $S_p$ can be approximated by a quadratic function (Okabe and Masuda 1984, Sadahiro 2001). Given this assumption, the formula that a critical point must satisfy is given (Takahashi et al. 1995),

$$\frac{\partial f}{\partial x} = \frac{\partial f}{\partial y} = 0 \tag{2}$$

with the corresponding three types of critical points being:

$$f = \begin{cases} -x^2 - y^2, & \text{peak (index = 2)} \\ x^2 + y^2, & \text{pit (index = 0)} \\ -x^2 + y^2, & \text{pass (index = 1)} \end{cases} \tag{3}$$

where the index indicates the number of negative eigenvalues of the Hessian matrix (denoted by $Hf$, Rana 2004):

$$Hf = \begin{pmatrix} f_{xx} & f_{xy} \\ f_{yx} & f_{yy} \end{pmatrix} = \begin{pmatrix} \frac{\partial^2 f}{\partial x^2} & \frac{\partial^2 f}{\partial x \, \partial y} \\ \frac{\partial^2 f}{\partial y \, \partial x} & \frac{\partial^2 f}{\partial y^2} \end{pmatrix} \tag{4}$$

where $f_{xx}$, $f_{xy}$, $f_{yx}$ and $f_{yy}$ are the partial derivatives of the function $f$. The basic three types of critical points are shown in Fig. 3.

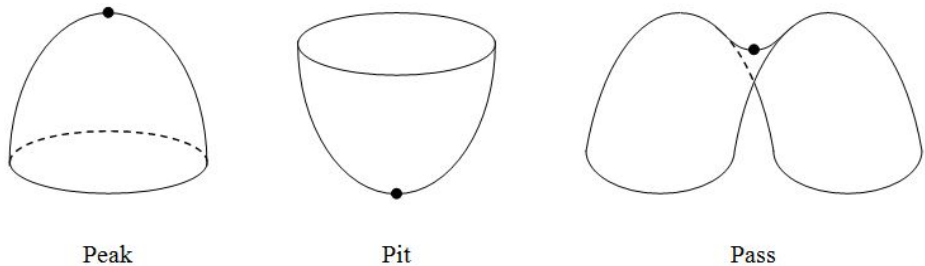

Peak       Pit       Pass

Figure 3. Three types of critical points (based on Okabe and Masuda 1984)

The critical points illustrated in Fig. 3 are non-degenerate with the matching Hessian matrix having a full rank. However, also possible is a degenerate critical point, known as a *monkey saddle* (see Fig. 4). Consequently, an extraction algorithm must be able to recognize and extract both basic and degenerate critical points. For degenerate critical points, we apply a decomposition to convert it into several non-degenerate common saddle points.

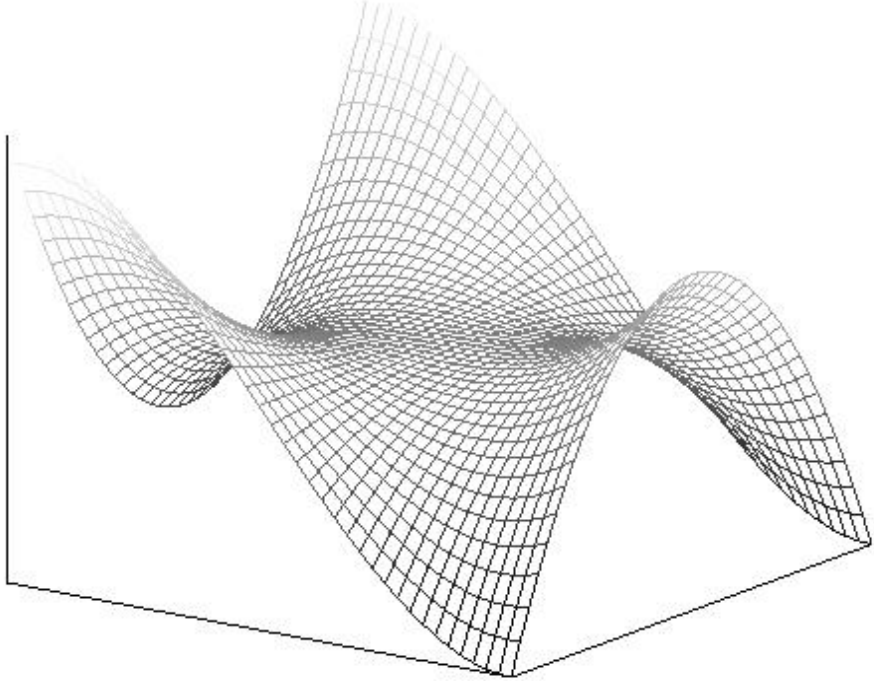

Figure 4. Monkey saddle

Conceptually, the KDE-generated surface is typically in raster format. However, Takahashi et al. (1995) claims that a raster elevation model lacks smoothness, affecting surface feature extraction. Therefore, we convert the raster to a *triangulated irregular network* (TIN) surface model to address this problem. In comparison with a raster model, a TIN only store fewer points to form their corresponding structures which can simplify the extraction process (Peucker et al. 1978). Additionally, the TIN is conceived of as a primary data structure which contains within it a secondary data structure in which the essential phenomena and features of the terrain (e.g. critical points and critical lines) are well-represented (Heil and Brych 1978, Mark 1978, Peucker et al. 1978).

**Algorithm 1: Critical feature extraction**

Input:   A topological surface $S = <P, Z>$, where $P = \{(x_1, y_1), (x_2, y_2), \dots , (x_n, y_n)\}$ representing a series of points consisting in the surface and $Z = \{z_1, z_2, \dots , z_n\}$ depicting corresponding height value associated with each point (e.g., $z_1$ represents the height value of point $P_1=(x_1, y_1)$)

Output:  Peaks, denoted by $G_{peak} = \{(x_1, y_1), (x_2, y_2), \dots , (x_k, y_k)\}$;

Pits, denoted by $G_{pit} = \{(x_1, y_1), (x_2, y_2), ... , (x_t, y_t)\}$;

Passes, denoted by $G_{pass} = \{(x_1, y_1), (x_2, y_2), ... , (x_s, y_s)\}$.

Step 1.   Let $GN = \{P_1, P_2, ...\}$ denote a collection of all neighbor points in counter-clockwise order with respect to the $xy$-coordinates around a point $P_m$ ($1 \leq m \leq n$), and $DN = \{(z_1 - z_m), (z_2 - z_m), ...\}$ represent a set of height differences between the point $P_m$ and its matching neighbor points.

Step 2.   Let $SN_i$ depict the corresponding positive or negative sign associated with each element in $DN$, and $N_s$ denote the total number of sign changes in the collection $SN = \{SN_1, SN_2, ... , SN_l\}$. (Note that, for completeness, we add $SN_1$ to the end of the $SN$ set.)

Step 3.   Find the first point $P_i$ ($1 \leq i \leq n$) in the point set $P$, and repeat the following procedure while $i \leq n$.

   a)   If the point $P_i$ has an identical height value with one of its neighbors $P_j$ (i.e. $z_i = z_j$), then:
   - If $x_i \neq x_j$, then let $z_i = x_i$ and $z_j = x_j$;
   - Otherwise, let $z_i = y_i$ and $z_j = y_j$.

   b)   Calculate corresponding values in $DN$, $SN$ and then $N_s$.

   c)   Let $N_{de}$ denote the number of passes after decomposition (e.g. a non-degenerate pass has a value of 1 while a degenerate pass has 2 or larger).

   d)   If $N_s = 0$, then do the following procedure:
   - If each element $SN_u$ in $SN$ satisfies $SN_u < 0$, then add the point $P_i$ into $G_{peak}$;
   - If each element $SN_u$ in $SN$ satisfies $SN_u > 0$, then add the point $P_i$ into $G_{pit}$.

   e)   Otherwise, let $N_{de} = (N_s$ - $2)/2$ and define a positive integer $v$ ($1 \leq v \leq N_{de}$), then repeat the following procedure until $v = N_{de}$:
   - Add the point $P_i$ into $G_{pass}$;
   - $v = v + 1$.

   f)   $i = i + 1$.

Step 4.   Exit.

The algorithm requires only $O(n)$ time in the worst case, where $n$ is the number of points in the point set $P$. It first detects the neighbors around each point and adds them into its neighbor collection in counter-clockwise order according to the $xy$-coordinates. The corresponding height difference between each neighbor point and the current target point is then measured and stored into a set. Note that a set of connected points might be tied due to limited precision of height values. Therefore, following Takahashi et al. (1995), we use the difference in x coordinates (or y coordinates if x coordinates are equal) as the height difference when height values of connected points are identical (see Step 3a). This is a rare case that requires some ordering for processing the points. Calculated next is the total amount of sign changes in the previous set (notably, this set starts and ends both with the height difference between the first neighbor point and the target point), resulting in the following classification:

(1). If all the height differences in the set are negative (i.e. the height of the current target point is higher than that of any matching neighbor points), the target point is labeled a peak.

(2). Conversely (i.e. all the height differences in the aforementioned set are positive), the target point is labeled a pit.

(3). Otherwise, the target point is labeled a pass if the number of sign changes is four or larger. For example, four sign changes of a target point implicitly indicate four critical lines (this term is explained below) of a common saddlepoint (specifically, the ridgeline and course-line are alternatively distributed around the saddlepoint). In addition, also possible is the mentioned degenerate pass such as monkey saddlepoint which has six critical lines, corresponding to six sign changes.

A degenerate pass (e.g. a monkey saddlepoint) may be encountered, especially with complex data such as mobile objects. We handle this case by decomposition to non-degenerate passes. First, we have to estimate how many non-degenerate passes exist. As shown in step 3e, the number of non-degenerate passes after decomposition (denoted by $N_{de}$) can be determined by the formula $(N_s - 2)/2$ where $N_s$ represents the total number of sign changes. After estimating the number of degenerate passes, we decompose a degenerate pass into corresponding set of non-degenerate ones. The details are explained below (see Fig. 6).

**Critical lines.** Critical lines correspond to surface features that connect critical points. A critical line is defined as follows (Takahashi et al. 1995). A curve on the topological surface is denoted by $C(t)$, it can be represented in the following form:

$$C(t) = \Big( C_x(t), C_y(t) \Big) \tag{5}$$

where $t$ ($t \in \mathbb{R}^+$) is a parameter. Suppose that the curve $C(t)$ satisfies the following differential equation:

$$\frac{\mathrm{d}C}{\mathrm{d}t} = -\Big( \frac{\partial C_x}{\partial t}, \frac{\partial C_y}{\partial t} \Big), C(0) = C_0 \tag{6}$$

where $C(0)$ is a point within the neighborhood of a pass $P$. The curve $C(t)$ is known as a ridge segment if it approaches the pass $P$ when $t$ approaches infinity. On the contrary, the curve $C(t)$ is known as a course segment if it satisfies the following equation:

$$\frac{\mathrm{d}C}{\mathrm{d}t} = \Big( \frac{\partial C_x}{\partial t}, \frac{\partial C_y}{\partial t} \Big), C(0) = C_0 \tag{7}$$

where $C(0)$ is still a point falling into the neighborhood of a pass $P$.

The critical line can be refined further. A *ridgeline* is either a single ridge segment or a chain of connected ridge segments, while a *course-line* is either a single course segment or a string of connected course segments. From the perspective of topological surface, a ridgeline ascends from a pass to a peak or another pass without reaching a pit (i.e. a ridgeline separates valleys), while a course-line descends from a pass to a pit or another pass without reaching a peak (i.e. a course-line separate hills) (Lee 1984). In the present study, however, we define a ridgeline as ascending from a pass to a peak, while a course-line as descending from a pass to a pit.

In accordance with the above definition of critical lines, a common saddlepoint has two ridgelines and two course-lines that lead to two corresponding peaks and two corresponding pits separately. In contrast, a degenerate pass has three corresponding ridgelines and course-lines. Therefore, to extract ridgelines and course-lines from a topological surface requires tracing all the slope lines associated with a pass that emanate from the pass to peaks or pits. In order to achieve the goal, we identify the four critical neighbors of a pass. For a non-degenerate pass, the total amount of sign changes in its matching *SN* set is four, in other words, there are exactly four places where the signs either change from positive to negative, or from negative to positive, in the *SN* set. For clarity, we define the terms *upper SN set* and *lower SN set* to separate different signs into these two lists. As regards the characteristic of a non-degenerate pass, all the elements within its *SN* set follow a pattern of alternate upper *SN* set and lower *SN* set. Furthermore, the number of upper *SN* set and that of lower one are identical, both equaling two. Figure 5 provides an illustration.

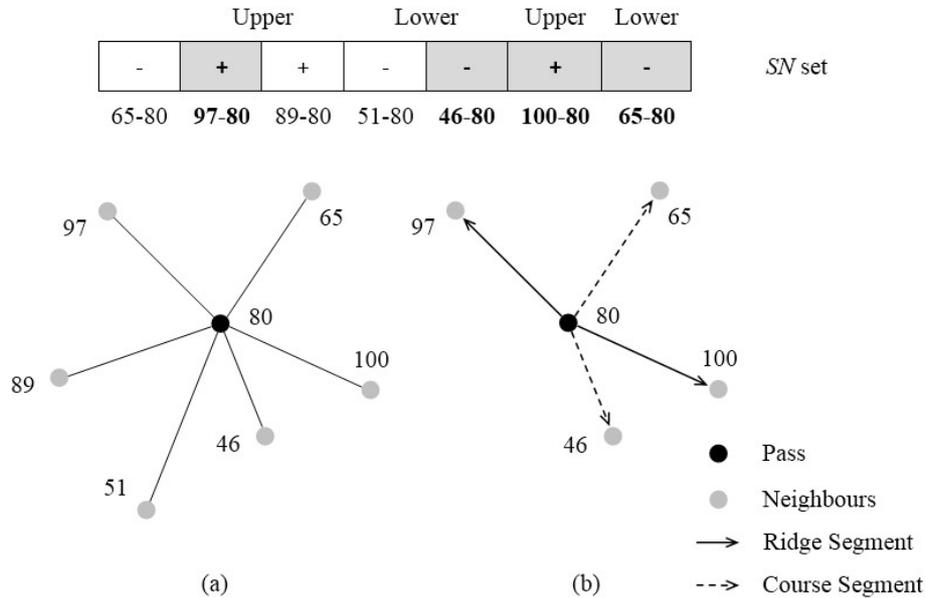

Figure 5. Exploring four critical neighbors of a non-degenerate pass
(a) All neighbors (b) Four critical neighbors

As shown in Fig. 5(b), the four neighbor points regarded as the four critical neighbors are the destination points of four steepest ascending and descending slope lines that emanate from the pass. Through tracing lines linking a pass and its four critical neighbors until reaching a peak or a pit, the ridgeline or the course-line associated with the pass is detected. Therefore, the extraction of critical lines is actually exploring the four critical neighbors of each pass. However, the process of detecting four critical neighbors associated with a degenerate pass is more complicated. Taking a monkey saddle as an example, all the upper and lower *SN* sets are first assigned separately. Afterwards, a procedure is conducted to simplify those sets, with only the element having the maximum height difference (in absolute value) remained in each upper and lower *SN*. The *GN* set of degenerate passes is then renewed in accordance with the neighbors that correspond to the elements in the upper and lower *SN* sets. For better assignment of the four critical neighbors to each degenerate pass, the *GN* set is then reordered, starting with the neighbor point whose height value is the lowest. With respect to a monkey saddle point, the $N_{de}$ (step 3e in the previous algorithm) is two, that is, it should be decomposed into two non-degenerate passes. Consequently, we first assign the first four neighbors in the renewed *GN* set to the first generated pass, and then assign the remaining points to another decomposed pass by removing the first two elements in the *GN* set. In this way, a degenerate pass is correctly decomposed into corresponding non-degenerate passes with their own four critical neighbors assigned. Figure 6 shows the related procedure.

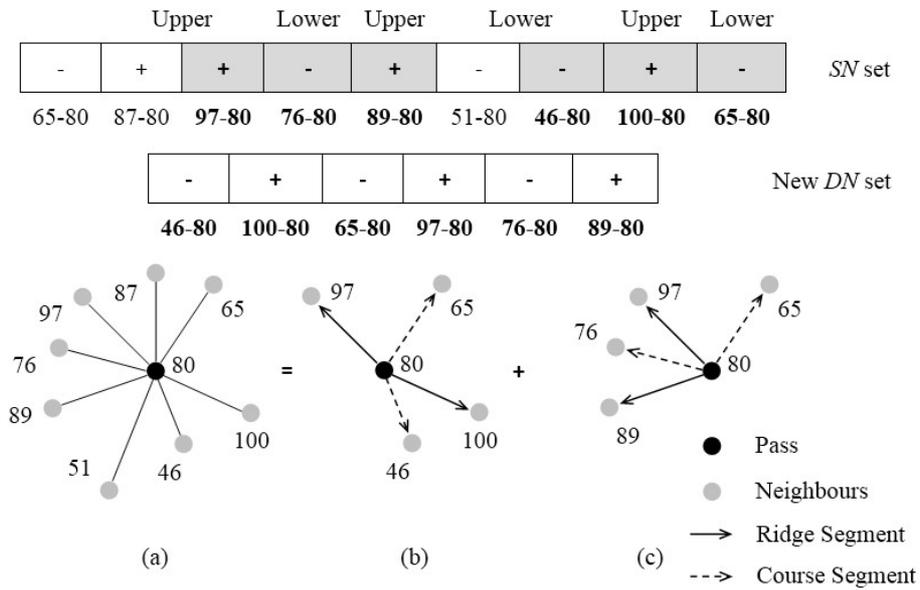

Figure 6. Exploring four critical neighbors of a degenerate pass
(a) All neighbors (b) Four critical neighbors of the first decomposed pass (c) Four critical neighbors of the second decomposed pass

Corresponding to the previous analyses, two functions are designed to find the four critical neighbors of a pass (non-degenerate or degenerate one) and two functions to detect the highest and lowest neighbors of a non-critical point, to simplify the representation of this algorithm below.

**Algorithm 2 − Critical lines extraction**

Function: *FindFourCriticalNeighbors_ n* ($P_m$), which is used for a non-degenerate pass;
 *FindFourCriticalNeighbors_ d* ($P_m$), which is used for a degenerate pass;
 *FindHighestNeighbor* ($P_m$) and *FindLowestNeighbor* ($P_m$), which are used for a non-critical point.

Input: A topological surface $S = <P, Z>$, where $P$ and $Z$ have the same meaning as that in the previous algorithm;
 The assigned peak set $G_{peak} = \{(x_1, y_1), (x_2, y_2), ... , (x_k, y_k)\}$;
 The assigned pit set $G_{pit} = \{(x_1, y_1), (x_2, y_2), ... , (x_t, y_t)\}$;
 The assigned pass set $G_{pass} = \{(x_1, y_1), (x_2, y_2), ... , (x_s, y_s)\}$.

Output: A ridgeline set $G_{ridgeline} = \{Sub\_ridgeline_{11}, Sub\_ridgeline_{12}, ... , Sub\_ridgeline_{s1}, Sub\_ridgeline_{s2}\}$;
 A course-line set $G_{courseline} = \{Sub\_courseline_{11}, Sub\_courseline_{12}, ... , Sub\_courseline_{s1}, Sub\_courseline_{s2}\}$, where $Sub\_ridgeline_{i1}$, $Sub\_ridgeline_{i2}$ and $Sub\_courseline_{i1}$, $Sub\_courseline_{i2}$ ($1 \leq i \leq s$) are the corresponding two ridgelines and two course-lines of the pass $P_i$ ($1 \leq i \leq s$).

Step 1. For each pass in $G_{pass}$, add an attribute $FCN = \{P_1, P_2, ...\}$ to store four critical neighbor points.

Step 2. Find the first pass point $P_i$ ($1 \leq i \leq s$) in $G_{pass}$ and repeat the following procedure while $i \leq s$:
 a) If $P_{i+1} \neq P_i$, then let $P_i.FCN = FindFourCriticalNeighbors\_n$ ($P_i$);
 b) Otherwise, let $P_i.FCN = FindFourCriticalNeighbors\_d$ ($P_i$) and $P_{i+1}.FCN = FindFourCriticalNeighbors\_d$ ($P_{i+1}$);
 c) $i = i + 1$.

Step 3. Find the first pass $P_j$ ($1 \leq j \leq s$) in $G_{pass}$, and then repeat the following procedure while $j \leq s$:
 a) Add $P_j$ to its matching ridgeline and course-line sets $Sub\_ridgeline_{j1}$, $Sub\_ridgeline_{j2}$ and $Sub\_courseline_{j1}$, $Sub\_courseline_{j2}$;
 b) Find the point $P_h$ with the highest height value in $P_j.FCN$, add it into $Sub\_ridgeline_{j1}$;
 c) Do the process below:
  ● If $P_h$ does not exist in $G_{peak}$, $G_{pass}$ or $G_{pit}$, then do the following procedure:
   ✓ Add $P_h$ into $Sub\_ridgeline_{j1}$;
   ✓ Let $P_h = FindHighestNeighbor$ ($P_h$), then go back to c);
  ● If $P_h$ is an element in $G_{pass}$, then go back to b);
  ● If $P_h$ is an element in $G_{peak}$, add $Sub\_ridgeline_{j1}$ into $G_{ridgeline}$, and then go to d);

d) Find the point $P_h$ with the second highest 'height' value in $P_j.FCN$, and add it into $Sub\_ridgeline_{j2}$;

e) Do the process below:
- If $P_h$ does not exist in $G_{peak}$, $G_{pass}$ or $G_{pit}$, then do the following procedure:
  - ✓ Add $P_h$ into $Sub\_ridgeline_{j2}$;
  - ✓ Let $P_h = FindHighestNeighbor\,(P_h)$, then go back to e);
- If $P_h$ is an element in $G_{pass}$, then go back to d);
- If $P_h$ is an element in $G_{peak}$, add $Sub\_ridgeline_{j2}$ into $G_{ridgeline}$, and then go to f);

f) Find the point $P_l$ with the lowest 'height' value in $P_j.FCN$, add it into $Sub\_courseline_{j1}$;

g) Do the process below:
- If $P_l$ does not exist in $G_{peak}$, $G_{pass}$ or $G_{pit}$, then do the following procedure:
  - ✓ Add $P_l$ into $Sub\_courseline_{j1}$;
  - ✓ Let $P_l = FindLowestNeighbor\,(P_l)$, then go back to g);
- If $P_l$ is an element in $G_{pass}$, then go back to f);
- If $P_l$ is an element in $G_{pit}$, add $Sub\_courseline_{j1}$ into $G_{courseline}$, and then go to h);

h) Find the point $P_l$ with the second lowest 'height' value in $P_j.FCN$, and add it into $Sub\_courseline_{j2}$;

i) Do the process below:
- If $P_l$ does not exist in $G_{peak}$, $G_{pass}$ or $G_{pit}$, then do the following procedure:
  - ✓ Add $P_l$ into $Sub\_courseline_{j2}$;
  - ✓ Let $P_l = FindLowestNeighbour\,(P_l)$, then go back to i);
- If $P_l$ is an element in $G_{pass}$, then go back to h);
- If $P_l$ is an element in $G_{pit}$, add $Sub\_courseline_{j2}$ into $G_{courseline}$, and then go to j);

j) $j = j + 1$.

Step 4. Exit.

As mentioned previously, this algorithm assigns four critical neighbors to each pass. Consequently, it first detects four critical neighbors for the two types of passes according to the processes shown in Fig. 5 and Fig. 6. Two ascending slope lines and two descending ones are then traced to explore the matching ridgelines and course-lines, until they reach two peaks and two pits, independently. Note that this algorithm requires $O(n)$ time, where $n$ is the number of elements in the assigned pass set $G_{pass}$.

### 3.3. Surface Network Measures

A surface network is a graph summarizing the structure of a topological surface. Given the focus on hotspot analysis in this paper, we construct our surface network using peaks and corresponding ridgelines. Peaks correspond to locations with relatively high mobility levels, and the connecting ridgelines allow us to focus on their spatial relationships and changes in these patterns with respect to time. In additional to visualizations of the peak and ridgeline network at each time period, we use graph indices to quantitatively summarize the network structure. As noted above, there is a large array of graph indices available; the appropriate index depends on the application question at hand. In this paper, we explore seven indices that cover a wide range of possible indices (Kansky and Danscoine 1989, Rodrigue et al. 2009):

**Cyclomatic Number (μ).** This measures the maximum number of independent cycles in a graph, indicating the complexity of a network:

$$\mu = e - v + p \tag{8}$$

where $e$, $v$ and $p$ are the number of edges (ridgelines), vertices (peaks) and sub-graphs contained in a surface network, respectively.

**Network Density (ND).** This measures the spatial coverage of a network. In the present case, this is the total kilometers of all ridgelines ($L$) per square kilometers of the surface area ($SA$) associated with the surface:

$$ND = L/SA \tag{9}$$

**Alpha (α).** This is a connectivity measure that evaluates the number of cycles in a network in comparison with the maximum number of cycles. This rages from zero to unity, with higher values indicting higher connectivity:

$$\alpha = \frac{e-v+p}{\frac{v(v-1)}{2}-(v-1)} \tag{10}$$

**Beta (β).** A measure of complexity, this is the ratio of edges to vertices in a network. High beta values indicate more complex structure:

$$\beta = e/v \tag{11}$$

**Gamma (γ).** A measure of connectivity that considers the relationship between the number of observed edges (ridgelines) and the number of possible edges in a network:

$$\gamma = \frac{e}{v(v-1)/2} \tag{12}$$

**Eta (η).** This is a measure of the relative spatial dispersion of the network; it is the average length per ridgeline in a surface network:

$$\eta = L/e \tag{13}$$

**Theta (ϑ).** Similar to eta, this is also a measure of the relative spatial dispersion of the network, but in this case relative to the number of peaks:

$$\vartheta = L/v \tag{14}$$

These seven indices can be further classified into two categories to conveniently measure and make comparisons among a series of surface networks. The cyclomatic number, network density and beta summarize the level of development and complexity of the network, while alpha, gamma, eta and theta measure network connectivity.

## 4. Example

We apply the surface network algorithms and measures described above to detect changes in patterns of mobile objects data in Shanghai, China. The mobile objects data consist of recorded GPS locations for about 2000 taxis from 12am to 6pm on a Friday, with a data collection interval of 20 seconds. The database consists of 43 million records with a storage size of about 3 GB. Attributes of each location point are listed in Table 1. The spatial extent of this data, as well as that of study area, is shown in Fig. 7. Note that the Yangtze River divides the study area into north and south subregions (and an island). Since few of the recorded taxi trips crossed the river we constructed independent surfaces for the north and south subregions and ignored the island.

Table 1. Attributes of a location point

| Car_ID | Time | Longitude | Latitude | Speed | Direction | IsLoad |
|--------|------|-----------|----------|-------|-----------|--------|
| ID of taxis | Recording time | Longitude of taxi's location | Latitude of taxi's location | Instantaneous speed of a taxi | Absolute direction relative to North | Whether a taxi is loaded |

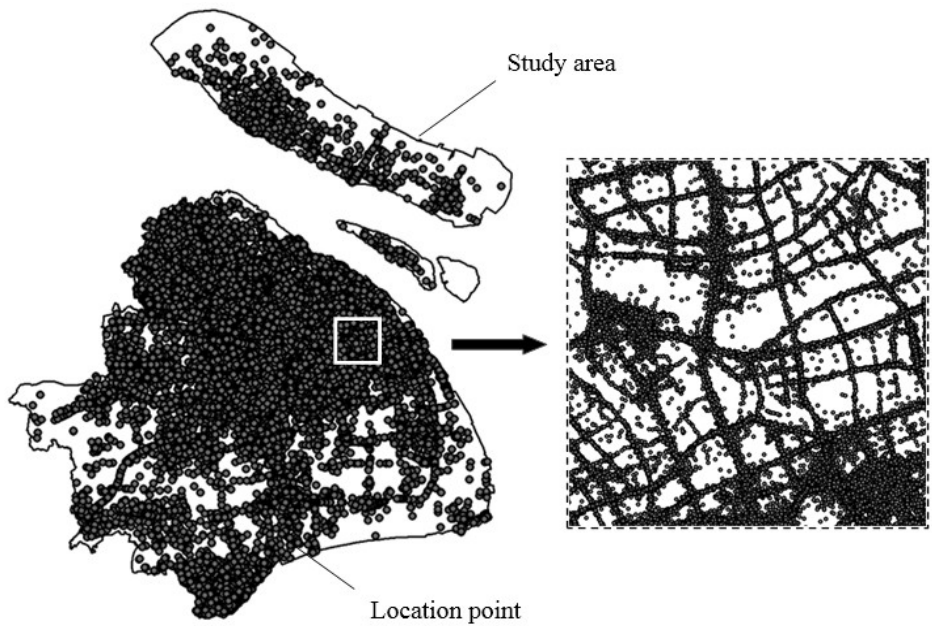

Figure 7. Map of study area illustrating recorded GPS points

Figure 8 illustrates the processing steps required to prepare the data for analysis. First, noisy data are detected and deleted. In this application, noisy data are taxi cab trips that have either or both ends locating beyond the study area due to GPS locating error (such as a record locating in the sea) or inter-province trips (such as trips from Shanghai to Jiangsu Province). Note that there are only few such records; and these are easily detected using spatial queries. The remaining data are then divided into 18 groups corresponding to hourly groupings (from 12am to 6pm - the time period for our data). We also extract the starting and ending points of each loaded (passenger) trip associated with each taxi. Since the original trajectory data are stored according to the receiving time at the server, the first ordering attribute of the data is not 'Car_ID' but 'Time'. Consequently, the data are first reordered in accordance with 'Car_ID'. All loaded trips of each taxi are then extracted based on the 'IsLoad' attribute. Finally, for each taxi, corresponding starting and ending points of each trip are recorded as an origin-destination matrix (OD matrix). In this way, the data volume is reduced by 98.1%, with roughly 820 thousand remaining location points.

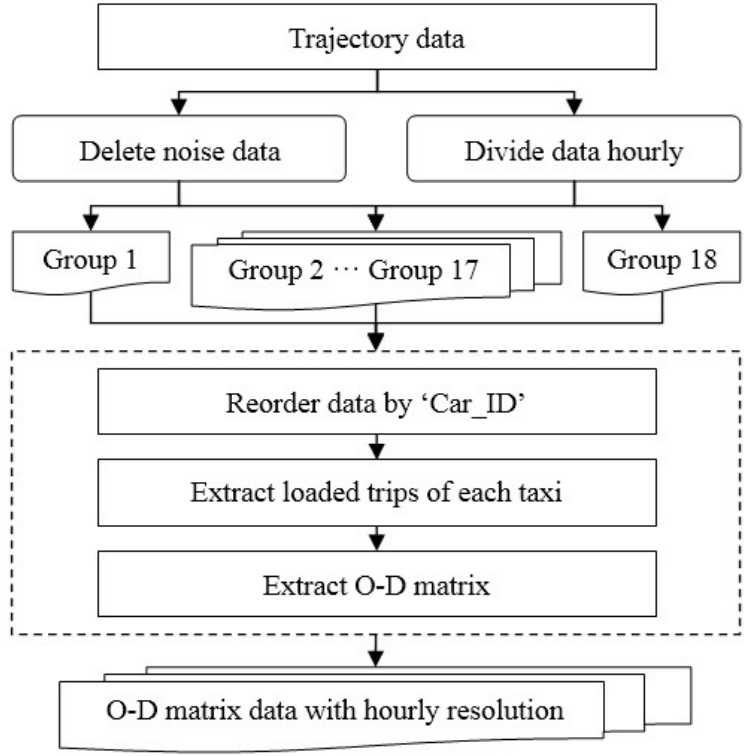

Figure 8. Data processing workflow

We experimented with different bandwidth settings to assess the effect on the resulting surface. Specifically, we applied the KDE method with bandwidth values corresponding to 150, 200, 600, 1250, 2000, 2871, where 2871 meters is the default bandwidth value suggested by ArcGIS based on the spatial resolution of the input data. Figures 9a-9c illustrate the resulting surfaces for bandwidths equal to 150, 600 and 2000 meters, respectively. A small bandwidth of 150 meters provides a surface with detail and noise that reflects the underlying street network, while a large bandwidth of 2000 meters generates a highly general and uninteresting surface. A bandwidth of 600 meters provides a reasonable balance between detail and generality; we use this surface to illustrate the remaining steps in the process.

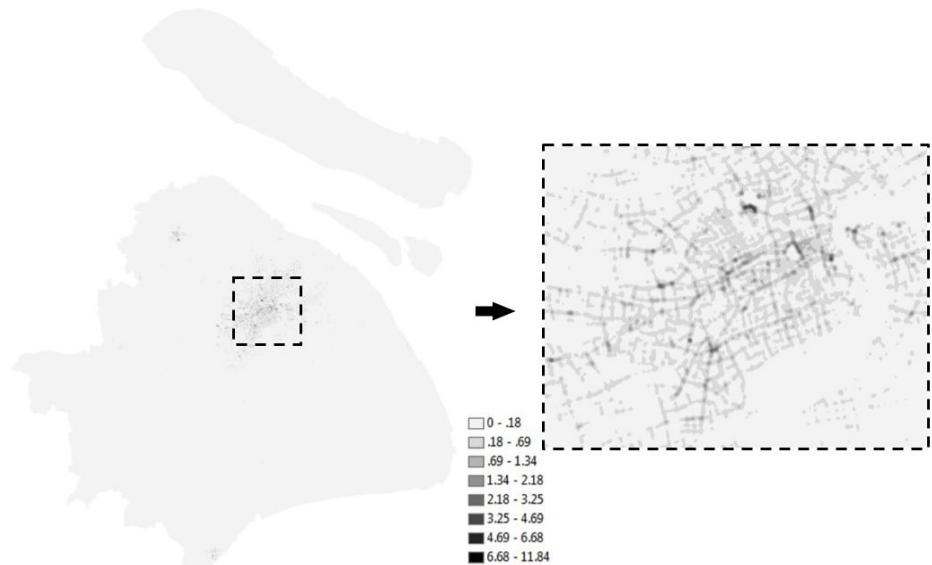

Figure 9 (a). Example generated surface with a KDE bandwidth of 150 meters

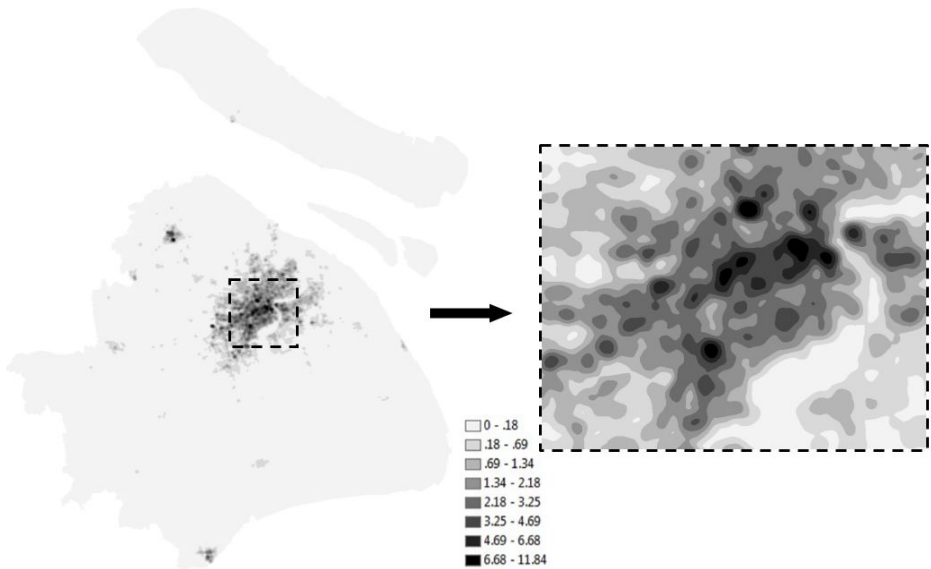

Figure 9 (b). Example generated surface with a KDE bandwidth of 600 meters

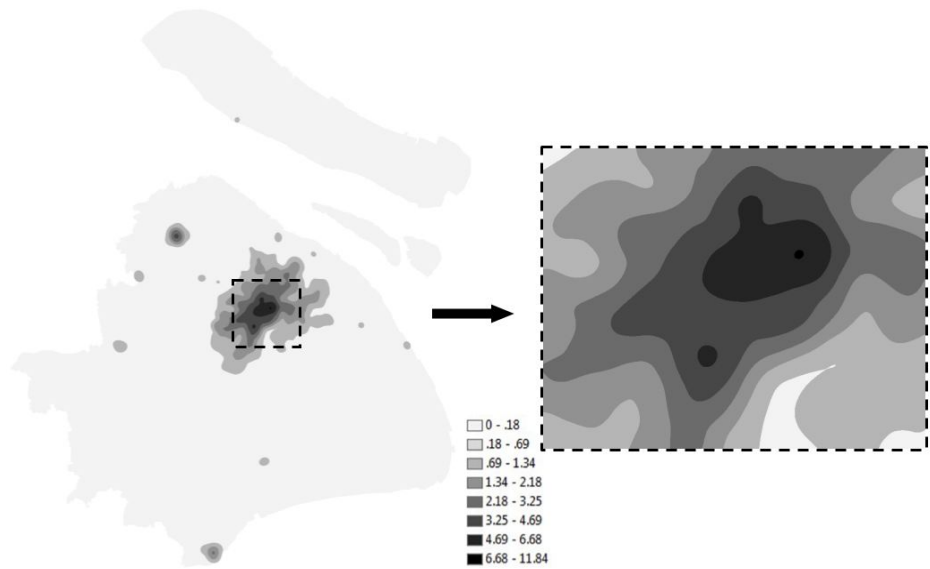

Figure 9 (c). Example generated surface with a KDE bandwidth of 2000 meters

We triangulate each generated surface into a TIN structure using the Delaunay Triangulation algorithm to extract critical points and critical lines. Figure 10 shows the 9am - 10am surface and corresponding triangulation for the 600 meter bandwidth surface.

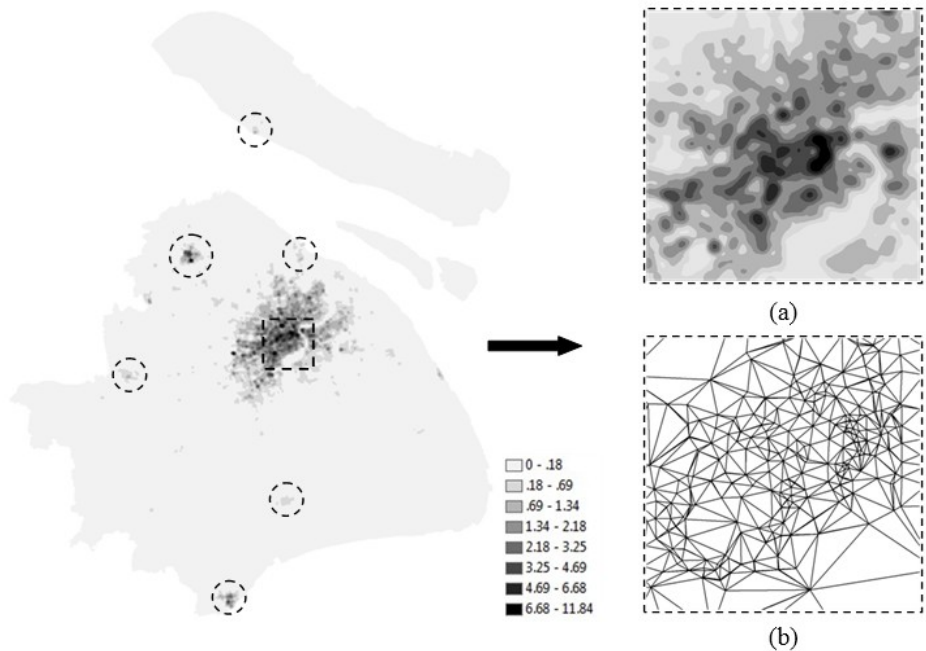

Figure 10. A surface and its corresponding triangulation. (a) Surface (b) Triangulation

In Fig. 10, regions with high mobility levels, i.e., hotspots, are highlighted in dark color. The most concentrated area - the central region of Shanghai - is marked with a dashed square. This area, consisting of 9 central districts of Shanghai (i.e. Jing'an, Huangpu, Xuhui, Changning, Hongkou, Yangpu, Putuo, Zhabei and Pudong New district), attracts the most people during this time period. Other than the central region, there are other locations frequently visited during this time period, indicated by six dashed circular regions. Interestingly, these six regions correspond perfectly to six administration areas within the city, matching intuition about hotspot distributions in Shanghai.

Table 2 summarizes the critical points extracted for each time period. The last column ("Peak+Pit-Pass") reports the summed extracted peaks and pits net the extracted passes. This relates to the Euler-Poincare formula that relates the number of vertices, edges and faces in a manifold (Okabe and Masuda 1984, Takahashi et al. 1995, Sadahiro 2001). According to this formula, the value should equal one for a smooth and continuous surface. The failure of these numbers to satisfy the Euler-Poincare formula is due to the discrete approximation of a continuous surface used in the computations. Jeong et al. (2014) prove that although the shape of a discretized surface may qualitatively capture the shape characteristic of a continuous surface, the number, location or existence of the critical points associated with a discretized surface can only approximate the critical points in the continuous surface. Therefore, the Euler-Poincare formula may not be satisfied.

Table 2. Summary of extracted critical points

| Time period (24 hour clock) | Peak | Pit | Pass | Peak+Pit-Pass |
|:---:|:---:|:---:|:---:|:---:|
| 0-1 | 68 | 49 | 114 | 3 |
| 1-2 | 61 | 38 | 97 | 2 |
| 2-3 | 35 | 17 | 50 | 2 |
| 3-4 | 41 | 23 | 62 | 2 |
| 4-5 | 42 | 22 | 62 | 2 |
| 5-6 | 34 | 15 | 47 | 2 |
| 6-7 | 24 | 16 | 38 | 2 |
| 7-8 | 39 | 20 | 56 | 3 |
| 8-9 | 58 | 33 | 88 | 3 |
| 9-10 | 91 | 51 | 139 | 3 |
| 10-11 | 86 | 58 | 141 | 3 |
| 11-12 | 66 | 40 | 103 | 3 |
| 12-13 | 68 | 32 | 98 | 2 |
| 13-14 | 68 | 36 | 102 | 2 |
| 14-15 | 66 | 41 | 104 | 3 |
| 15-16 | 66 | 46 | 109 | 3 |
| 16-17 | 65 | 41 | 103 | 3 |

| | | | |
|---|---|---|---|
| 17-18 | 72 | 44 | 113 | 3 |

Note that the method may detect peaks even if the height difference between a location and its neighbor is very small. Therefore, to reduce minor, superfluous peaks, we apply a relative threshold for peaks; requiring a 10% difference in height relative to their surrounding neighborhood for peaks to remain in the analysis (Sadahiro, 2001). Table 3 summarizes the results. Comparing the number of peaks in Table 2 and Table 3, note that several insignificant peaks have been deleted.

Table 3. Significant peaks for each hourly group

| Hour | Peaks | Hour | Peaks |
|---|---|---|---|
| 0-1 | 51 | 9-10 | 71 |
| 1-2 | 43 | 10-11 | 66 |
| 2-3 | 24 | 11-12 | 51 |
| 3-4 | 23 | 12-13 | 50 |
| 4-5 | 22 | 13-14 | 55 |
| 5-6 | 17 | 14-15 | 52 |
| 6-7 | 10 | 15-16 | 52 |
| 7-8 | 21 | 16-17 | 52 |
| 8-9 | 46 | 17-18 | 56 |

The results in Table 3 suggest that the number of hot spots fluctuates with time. They decrease from 12am to 6am when the number is at the lowest and then rise sharply until 9am when they reach their greatest number. The number of hot spots then decreases sharply until 12pm-1pm and then remains flat until the end of the time horizon (6pm). This general pattern fits well with our intuition about daily activity patterns in the study city. Figure 11 illustrates one of the 18 surface networks generated in our analysis (9am-10am; the same as Fig. 10). Note that all surface networks illustrated here and after are composed of only significant peaks and corresponding ridgelines.

As noted above, we chose 600 meters as the appropriate bandwidth in the KDE method. To confirm this bandwidth choice, we also applied the significant peak threshold test and extracted ridgelines for the corresponding surfaces generated with bandwidths of 150 and 2000 meters. Figure 12 and 13 illustrate the surface networks extracted for these KDE bandwidth values, respectively. Figure 12 is not a satisfactory representation of the hot spot spatial distribution: similar to an overtrained neural network, the surface network matches too closely the specific context of the study area, namely, the underlying street network. Conversely, the surface network in Figure 13 does not tell us much about the distribution of high mobility locations in the study area.

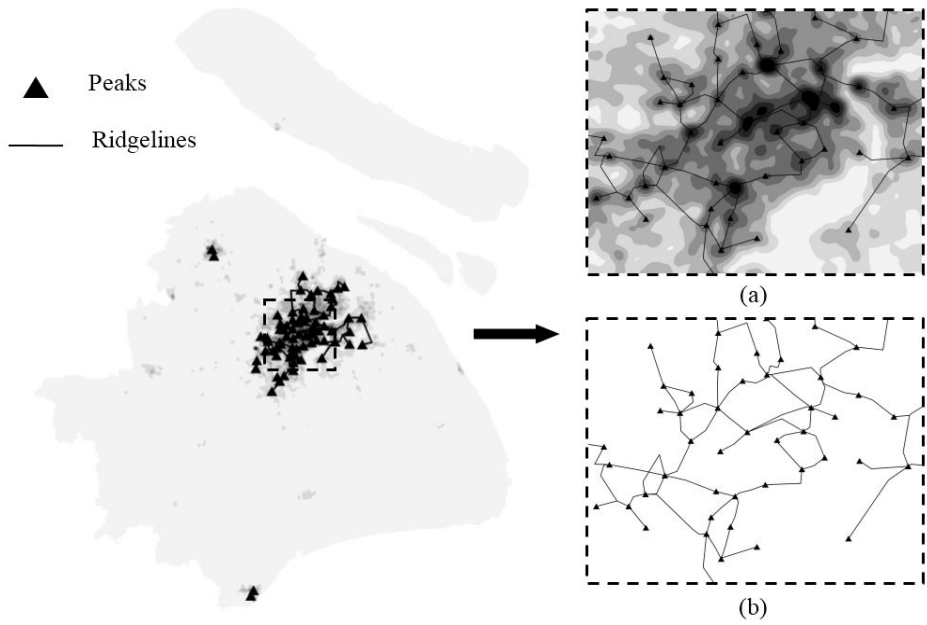

Figure 11. Example surface network extracted from a 600m bandwidth KDE resulted surface

(a) Details of surface and corresponding surface network (b) Surface network alone

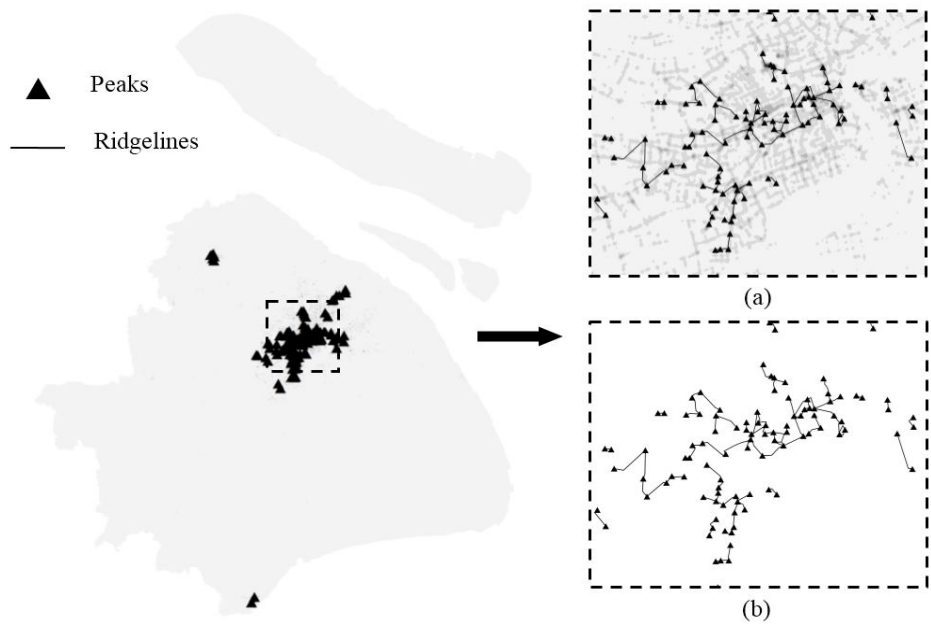

Figure 12. Example surface network extracted from a 150m bandwidth KDE resulted surface

(a) Details of surface and corresponding surface network (b) Surface network alone

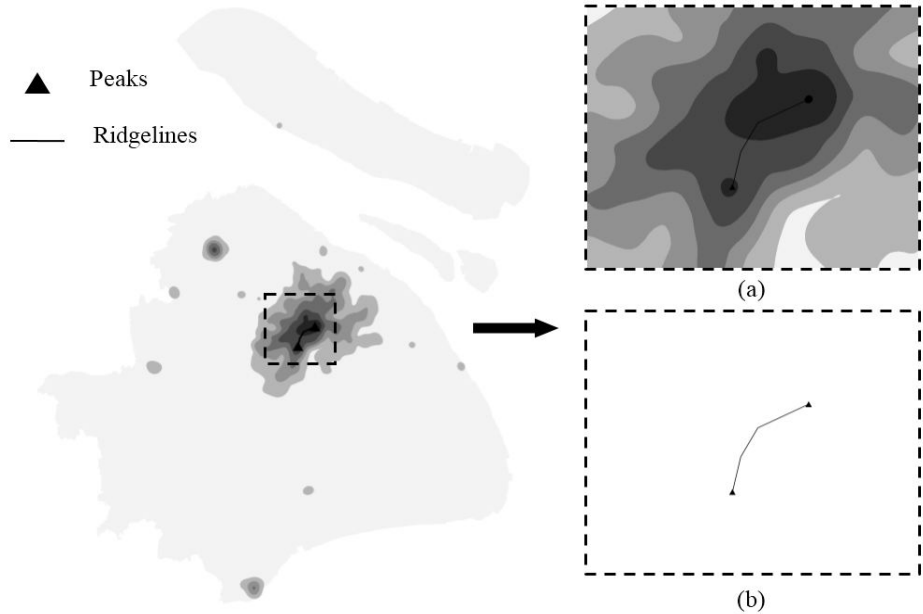

Figure 13. Example surface network extracted from a 2000m bandwidth KDE resulted surface

(a) Details of surface and corresponding surface network (b) Surface network alone

Finally, we apply the seven graph indices to assess changes in the hotspot spatial patterns over time. Figure 14 shows the results of indices representing the development and complexity of the surface networks (cyclomatic number, network density and beta). From 12am to 5am, these three indices all decrease in value, indicating that the surface network becomes less complex as the number of mobility hotspots decreases and their spatial distribution shrinks. This trend continues until about 6am when they reach the lowest point; for example, the cyclomatic number drops to zero between 5am and 6am, indicating that the spatial distribution pattern of hotspots is the simplest and sparsest. However, after 6am, the complexity indicators start to rise, with a small peak emerging between 6am and 7am. From 7am they decrease slightly until 8am. A plausible explanation is that people who live far away from their work places must start out early in the morning, usually between 6am and 7am. Since they live far from work, the taxis fees are high. Therefore, many people may take taxis to the closest metro stations near their living places. People who live less distant from their workplaces travel between 7am and 8am, with many heading directly to their workplaces via a taxi. As work time in Shanghai begins at 9am, the corresponding number of hot spots during this time reaches another peak with a high development and complexity of surface network. After work starts, the number of people taking taxis begins to decrease, and the surface network decreases slightly in its complexity but remains relatively high. The fewer taxi rides during this period being for diffuse purposes such as shopping, friends and errands, leading to a fairly high degree to complexity in the surface network describing the spatial distribution of the hotspots. Overall, it appears that mobility patterns shift from simple patterns during the night to more complex patterns as people awake and conduct their daily activities. The findings correspond well to the human activities such as going to work and going to lunch on weekdays in Shanghai (Liu et al. 2012).

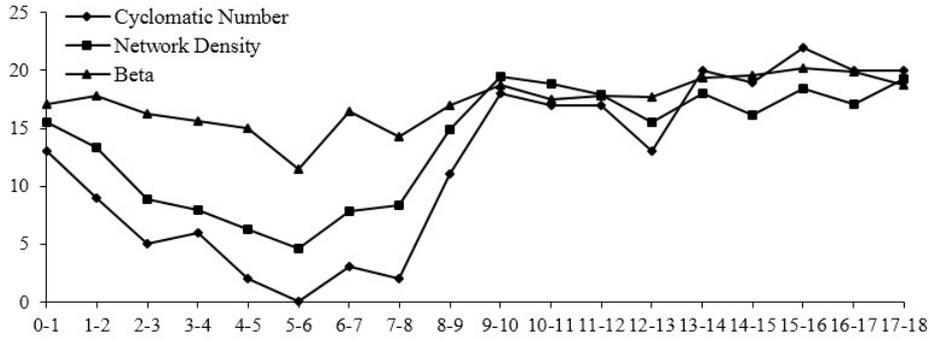

Figure 14. Graph indices over time: Complexity measures

Figure 15 illustrates the behavior of the connectivity indices over time. The surface network describing the spatial pattern of mobility hotpots generally suggests low connectivity, with a substantial increase in connectivity during the 6am-7am time period. These results suggest an interaction between the number of hotspots, their spatial concentration and the connectivity of the resulting network. Note that during the 12am-5am periods the number and complexity of hotspots becomes lower (see Fig. 14), but their connectivity starts to rise during this period. As people travel less and the number of hotspots decline, taxi drivers frequently concentrate on well-known locations in small regions where the probability of passengers is highest. The dramatic increase hotspot connectivity between 6am and 7am is also consistent with this explanation: distant workers focus their demand for rides between home and local metro stations, meaning that the few hotspots will be densely (but locally) concentrated. After the 8am-9am period, all indices indicate low connectivity; this can be explained by more diffuse demand for taxis across a wider range of activities and locations.

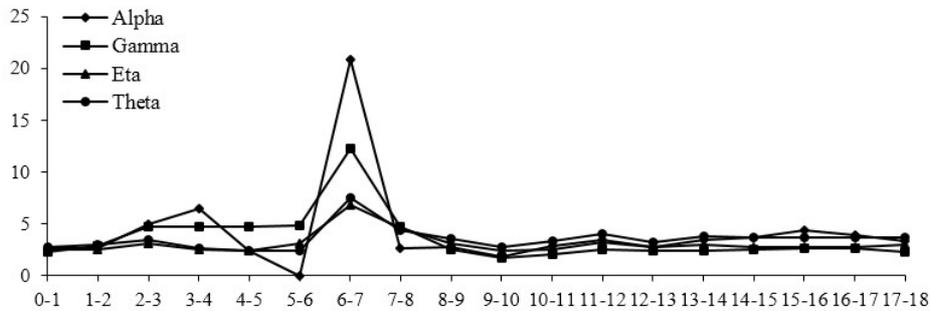

Figure 15. Graph indices over time: Connectivity measures

## 5. Conclusion

Summarizing large collections of mobile objects is difficult given the dramatic increasing capabilities for collecting and storing data on mobile objects over the past few decades. This paper develops methods for identifying and analyzing hotspots in collections of mobile objects. We use kernel density estimation (KDE) to summarize the locations of mobile objects over some interval of time as a smooth, continuous surface, develop algorithms to simplify the surface by extracting critical points (peaks, pits and passes) and critical lines (ridgelines and course-lines) and connect the peaks and ridgelines to produce a surface network. We apply graph theoretic measures to changes in the surface networks representing mobile objects over a sequence of time periods. To illustrate our approach, we apply the techniques to taxi cab data collected in Shanghai, China. The findings match well with scientific and anecdotal knowledge about human activity patterns in the study area.

Next steps in this research include more fully parameter search and sensitivity analysis for the KDE method and evaluating the representativeness of the hotspots using auxiliary data such as traffic counts and behavioral surveys of taxi drivers. In addition, we chose a wide range of graph theoretic measures for exploration and illustration purposes; a careful evaluation of these and other measures regarding their semantics and suitability with respect to mobility hotspots is warranted. Finally, developing and applying measures that examine network topology dynamics more directly could be useful for gaining insights into hotspot dynamics.

## Acknowledgments

This research is supported by the National High Technology Research and Development Program of China (No. 2013AA122302), Innovation Research Grant of East China Normal University (No. 239201278220056), Shanghai Natural Science Foundation (No. 11ZR1410100) and National Natural Science Foundation of China (No. 41271441). We also thank the three anonymous referees for their valuable comments.